%% file: main.tex
\documentclass[10pt,twocolumn,letterpaper]{article}

\usepackage{cvpr}              % To produce the CAMERA-READY version
\newif\ifincludeappendix
\includeappendixtrue


\usepackage{cuted}
\usepackage{multirow} 
\usepackage{balance}
\definecolor{cvprblue}{rgb}{0.21,0.49,0.74}
\usepackage[pagebackref,breaklinks,colorlinks,allcolors=cvprblue]{hyperref}

\newcommand{\work}{\textit{DSplats}}

\def\paperID{2152} % *** Enter the Paper ID here
\def\confName{CVPR}
\def\confYear{2025}

\title{DSplats: 3D Generation by Denoising Splats-Based Multiview Diffusion Models}

\author{Kevin Miao \\
Apple\\
{\tt\small k\_miao@apple.com}
\and
Harsh Agrawal \\
Apple \\
{\tt\small hagrawal2@apple.com}
\and
Qihang Zhang \\
CUHK, Apple \\
{\tt\small qhzhang@link.cuhk.edu.hk}
\and
Federico Semeraro \\
Apple \\
{\tt\small f\_semeraro@apple.com}
\and
Marco Cavallo \\
Apple \\
{\tt\small marco\_cavallo@apple.com}
\and
Jiatao Gu \\
Apple \\
{\tt\small jiatao@apple.com}
\and
Alexander Toshev \\
Apple \\
{\tt\small toshev@apple.com}
}

\begin{document}
\maketitle
\input{Paragraphs/0-abstract.tex}
\input{Paragraphs/1-introduction.tex}
\input{Paragraphs/2-related.tex}
\input{Paragraphs/3-method.tex}
\input{Paragraphs/4-experiments.tex}

\input{Paragraphs/5-conclusion.tex}
{
    \small
    \balance
    \bibliographystyle{ieeenat_fullname}
    \bibliography{main}
}

\input{Paragraphs/6-appendix}

\end{document}

%% file: Paragraphs/0-abstract.tex
\begin{strip}
\centering
\includegraphics[width=0.8\textwidth]{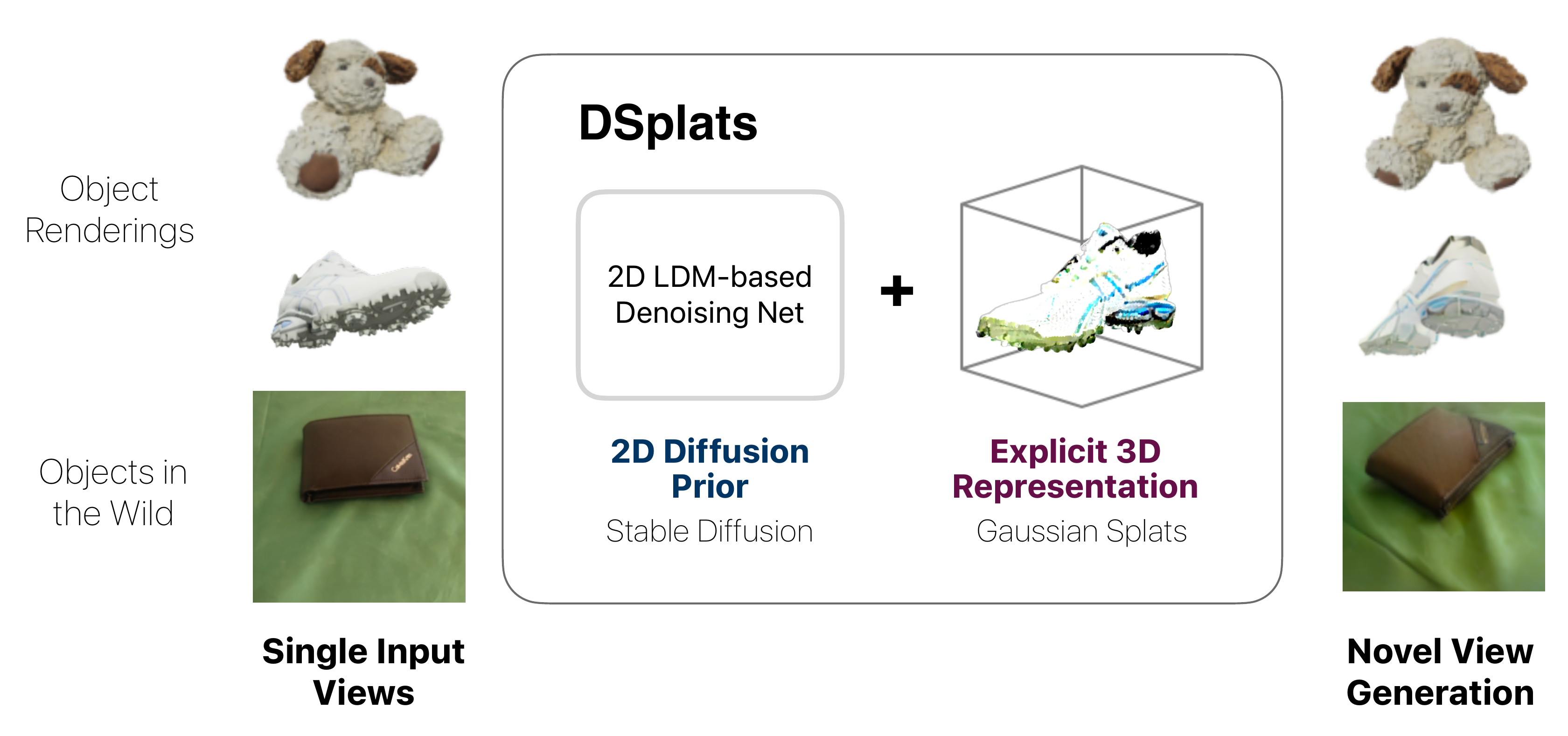}
\captionof{figure}{By leveraging the 2D Diffusion Prior of Latent Diffusion Models and an explicit 3D Gaussian representation, \work{} is able to generate photorealistic 3D objects when provided with a single image input only. These objects can then be rendered from any novel view, including objects in the wild.}
\label{fig:intro}
\end{strip}

\begin{abstract}
Generating high-quality 3D content requires models capable of learning robust distributions of complex scenes and the real-world objects within them. Recent Gaussian-based 3D reconstruction techniques have achieved impressive results in recovering high-fidelity 3D assets from sparse input images by predicting 3D Gaussians in a feed-forward manner. However, these techniques often lack the extensive priors and expressiveness offered by Diffusion Models. On the other hand, 2D Diffusion Models, which have been successfully applied to denoise multiview images, show potential for generating a wide range of photorealistic 3D outputs but still fall short on explicit 3D priors and consistency. In this work, we aim to bridge these two approaches by introducing \work, a novel method that directly denoises multiview images using Gaussian Splat-based Reconstructors to produce a diverse array of realistic 3D assets. To harness the extensive priors of 2D Diffusion Models, we incorporate a pretrained Latent Diffusion Model into the reconstructor backbone to predict a set of 3D Gaussians. Additionally, the explicit 3D representation embedded in the denoising network provides a strong inductive bias, ensuring geometrically consistent novel view generation. Our qualitative and quantitative experiments demonstrate that \work{} not only produces high-quality, spatially consistent outputs, but also sets a new standard in single-image to 3D reconstruction. When evaluated on the Google Scanned Objects dataset, \work{} achieves a PSNR of 20.38, an SSIM of 0.842, and an LPIPS of 0.109.
\end{abstract}

%% file: Paragraphs/1-introduction.tex
\section{Introduction}

    The demand for generating controllable, high-quality 3D objects and scenes is growing rapidly across industries like spatial computing, robotics, gaming, motion pictures, architecture and healthcare. As these fields push toward creating more realistic simulations, immersive experiences and interactive environments, there is an ever-growing need for scalable 3D generation methods that can keep up with this demand. In contrast with 2D and video generation methods, there are several unique key challenges to 3D content generation. The vast amount of online images and video content far exceeds the available 3D assets and scenes by several orders of magnitude. While recent initiatives have greatly increased the number of 3D datasets~\citep{objaverse, objaverseXL}, the available data contains many samples that are either low-quality or differ from the distribution of real-world objects, which are precisely the types of assets that artists and developers often need to either generate or work with. 
    
    One approach in this field focuses on learning neural 3D representations of scenes or objects from a set of images with corresponding viewpoints. These neural representations can be used to render novel views from arbitrary angles or extract textured meshes~\citep{guedon2024sugar, mildenhall2021nerf}. More recently, Gaussian Splatting (3DGS)~\citep{kerbl20233d} has emerged as a powerful method, characterized by an explicit 3D Gaussian representation. 3DGS achieves faster optimization times and demonstrates the capacity to capture high levels of detail, even for extensive scenes~\citep{lin2024vastgaussian}. However, these methods still depend on a large number of clean input views to produce high-quality novel perspectives.

Large reconstruction models, such as LRM~\citep{hong2023lrm}, have addressed this limitation by enabling 3D reconstruction with a sparse set of views, effectively making the process more efficient~\citep{szymanowicz2024splatter, tang2024lgm}. Although these models reduce data requirements, they still face challenges in achieving fine-grained detail and expressiveness, primarily due to their deterministic nature.

In parallel, another line of research leverages the rich priors of 2D representations and video generative models to generate 3D assets~\citep{poole2022dreamfusion, wang2023imagedream}. These methods use 2D diffusion models to construct 3D-consistent multiview images. Despite their innovative approach, they remain limited in terms of quality, optimization speed, and the inability to directly generate 3D representations.
    
In this paper, we introduce a 3D diffusion model, named \work{}, that combines the strengths of two key approaches: the expressive and rich prior of image diffusion models~\cite{rombach2022high} and the explicit 3D modeling capabilities of Gaussian Splat-based reconstructors~\citep{kerbl20233d} (see Fig.~\ref{fig:intro}).

Specifically, \work{} learns a Latent Diffusion Model that operates simultaneously on multiple views of an object. This model denoises a set of latents for these views within a single differentiable network, executed in two main steps. First, it maps the latents to a 3D Gaussian representation of the object. Then, it renders these Gaussians and re-encodes them into latents. During training, \work{} learns to denoise all latents corresponding to multiple views of the same object using a consistent 3D representation. At inference, the model can generate either an explicit 3D model or novel views directly from a single input view.

\work{} integrates two complementary submodels that are mutually beneficial. The first submodel initializes the latents-to-3D-Gaussians network using a Latent Diffusion Model pre-trained on a large collection of 2D images~\cite{rombach2022high}, leveraging the extensive prior knowledge embedded in 2D generative models. The second submodel introduces an explicit 3D representation as an intermediate activation within the diffusion process, serving as a natural inductive bias to enforce consistency across latents for different views of the same object. During training, this 3D representation enables an image consistency loss that guides the denoising model to generate views closely resembling real ones. Leveraging the differentiability of 3D Gaussian Splatting, end-to-end training is achieved seamlessly. During inference, this approach facilitates the direct rendering of novel views that maintain consistency with both the input view and one another.

We evaluate \work{} extensively on the Google Scanned Objects dataset~\cite{downs2022google}, achieving state-of-the-art results across multiple metrics. The generated novel views exhibit both high visual realism and strong geometric consistency. We attribute the former to the 2D diffusion model prior, while the latter is strengthened by the explicit 3D representation.

%% file: Paragraphs/2-related.tex
\section{Related Works}

\begin{figure*}
    \centering
    \includegraphics[width=1.0\linewidth]{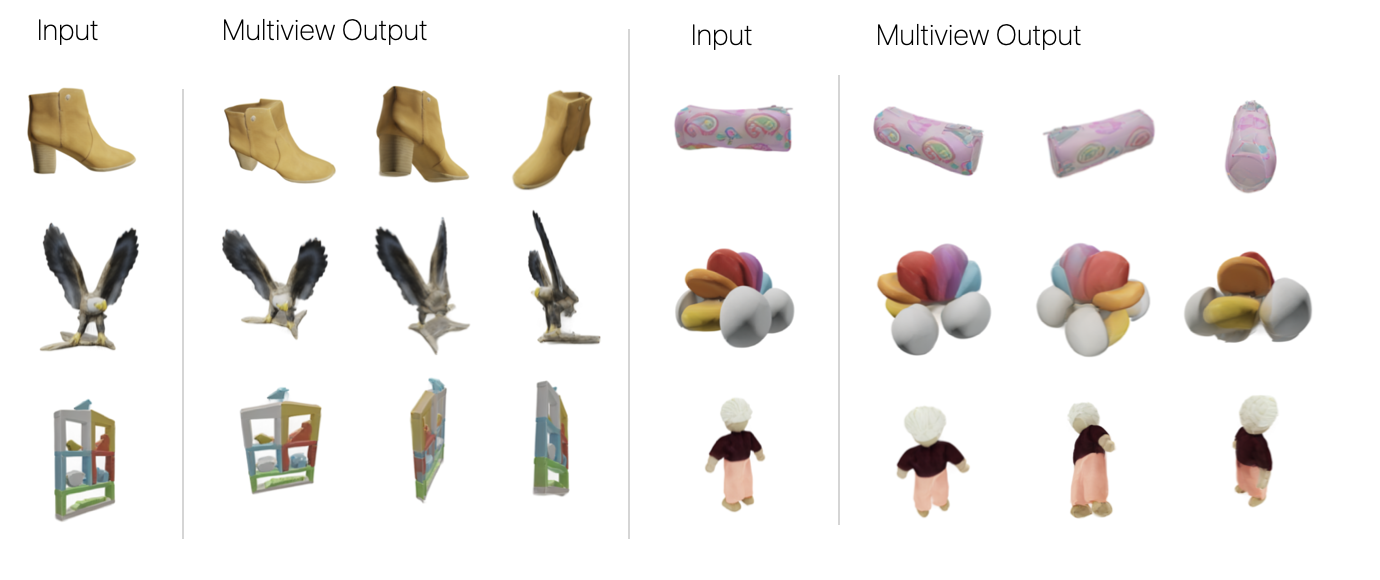}
    \caption{Qualitative results: provided a single input image of real-world objects, \work{} is able to generate high-quality 3D representations, yielding realistic 3D objects.}
    \label{fig:results}
\end{figure*}

The generation of high-quality 3D content from sparse 2D views is a challenging problem that has been explored across various domains, including neural rendering, generative models, and diffusion techniques. Here, we discuss relevant prior work in 3D reconstruction, diffusion models, and reconstructor-based denoisers that have informed our approach.

\textbf{3D Representations}
\textbf{Sparse-View Reconstruction.}
Early approaches to 3D scene representation and novel-view synthesis (NVS) have laid the foundation for sparse-view reconstruction methods. Techniques such as NeRF \cite{mildenhall2021nerf}, 3DGS \cite{kerbl20233d}, and Neural Graphics Primitives (NGP) \cite{muller2022instant} have successfully rendered high-quality 3D scenes for view interpolation. However, these methods typically require dense multi-view images to produce photorealistic results, which limits their application in sparse-view settings.

For sparse-view reconstructions, models using Score Distillation Sampling (SDS) \cite{poole2022dreamfusion} introduced methods for lifting 2D priors into 3D. These include Point-E \cite{nichol2022point}, DreamFusion \cite{poole2022dreamfusion}, and ImageDream \cite{wang2023imagedream}, which combine 2D diffusion models with differentiable rendering. Multiview diffusion models address the challenge of generating consistent multiview images by leveraging pretrained 2D Diffusion Models, further conditioned on camera parameters. Unlike optimization methods like SDS, multiview diffusion models can directly predict spatially consistent multiview images, significantly reducing inference time while benefiting from the large 2D prior of the diffusion model. These multiview diffusion models still have several limitations related to the quality, optimization time, camera-control and view-consistency. Additionally, to generate novel views or extract actual 3D meshes, these methods still require a second training step converting these views into a 3D representation, thus involving a NeRF or 3DGS method.

To improve efficiency and scalability, triplane-based methods like LRM \cite{hong2023lrm}, LRM-Zero \cite{xie2024lrm}, MeshLRM \cite{wei2024meshlrm}, and TripoSR \cite{tochilkin2024triposr} introduced triplane representations to sparse-view 3D reconstruction. For example, at inference time, \citet{hong2023lrm} uses a transformer model to predict triplane features given a single or sparse set of images along with camera ray maps. These triplane features are subsequently used to train a NeRF model. This and above methods optimize reconstruction quality and generalization to unseen views while balancing memory and speed.

Recent advances in explicit representations of 3D objects have explored Gaussian-based models such as LGM \cite{tang2024lgm}, SplatterImage \cite{szymanowicz2024splatter}, and GRM \cite{xu2024grm}, which use 3D Gaussian splats to capture object shapes in feed-forward setups. These models provide fast inference times and scalable generalization, as seen in Instant3D \cite{li2023instant3d}, which also extends single-view conditioning to multiple views. By predicting 3D Gaussians directly, our work builds on these efficient methods while introducing a 2D diffusion model prior to improve visual fidelity and view consistency.

\textbf{Diffusion Models for 3D Generation.}
Diffusion models have emerged as a powerful tool for 2D image generation and, more recently, for multiview and 3D content generation. For multiview diffusion, methods like Zero123++ \cite{shi2023zero123++}, One-2-3-45++ \cite{liu2024one}, MVDream \cite{shi2023mvdream}, and MVDiffusion \cite{tang2023emergent} have applied denoising models to generate spatially consistent images, conditioning on camera poses to generate novel views without 3D structure.

In the domain of video generation and temporally coherent diffusion, models like SV3D \cite{voleti2025sv3d} and SV4D \cite{xie2024sv4d} have explored applying diffusion to sequential 3D views, maintaining temporal coherence in multiview synthesis. However, such models are often limited in generalizing to complex, high-fidelity scenes.

Pose-conditioned diffusion models like CAT3D \cite{gao2024cat3d}, ReconFusion \cite{wu2024reconfusion}, and ZeroNVS \cite{sargent2023zeronvs} introduce pose-awareness into 2D diffusion models, enhancing view consistency by learning camera-conditioned image distributions. By integrating 2D latent diffusion within a 3D Gaussian framework, our model leverages the rich priors of 2D diffusion models, enhancing quality in a multiview context.

\textbf{Reconstructor-Based Denoisers.}
Reconstructor-based denoisers have emerged as a promising approach for view-consistent 3D content generation, with methods like DMV3D \cite{xu2023dmv3d}, Viewset Diffusion \cite{szymanowicz2023viewset}, and RenderDiffusion \cite{anciukevivcius2023renderdiffusion} demonstrating success in multiview image denoising. These models employ 3D reconstructor backbones for latent space denoising, significantly reducing inference time and improving view consistency across generated images.

Building on these approaches, our model directly incorporates a 3D Gaussian reconstructor as the denoising mechanism within a latent diffusion model. This integration enables efficient denoising of multiview images while utilizing large-scale 2D priors, avoiding the need for time-consuming optimization steps associated with SDS methods. Furthermore, our model's use of Gaussian Splatting allows for high-fidelity detail while preserving spatial consistency across views. While previous work by~\citet{chen3d} successfully integrated pretrained latent diffusion models with Gaussian reconstruction models, it still requires a two-step training approach. In \work{}, the diffusion and reconstruction training occur in a single-stage.

%% file: Paragraphs/3-method.tex
\section{Method}
\begin{figure*}[ht]
    \centering
\includegraphics[width=\textwidth]{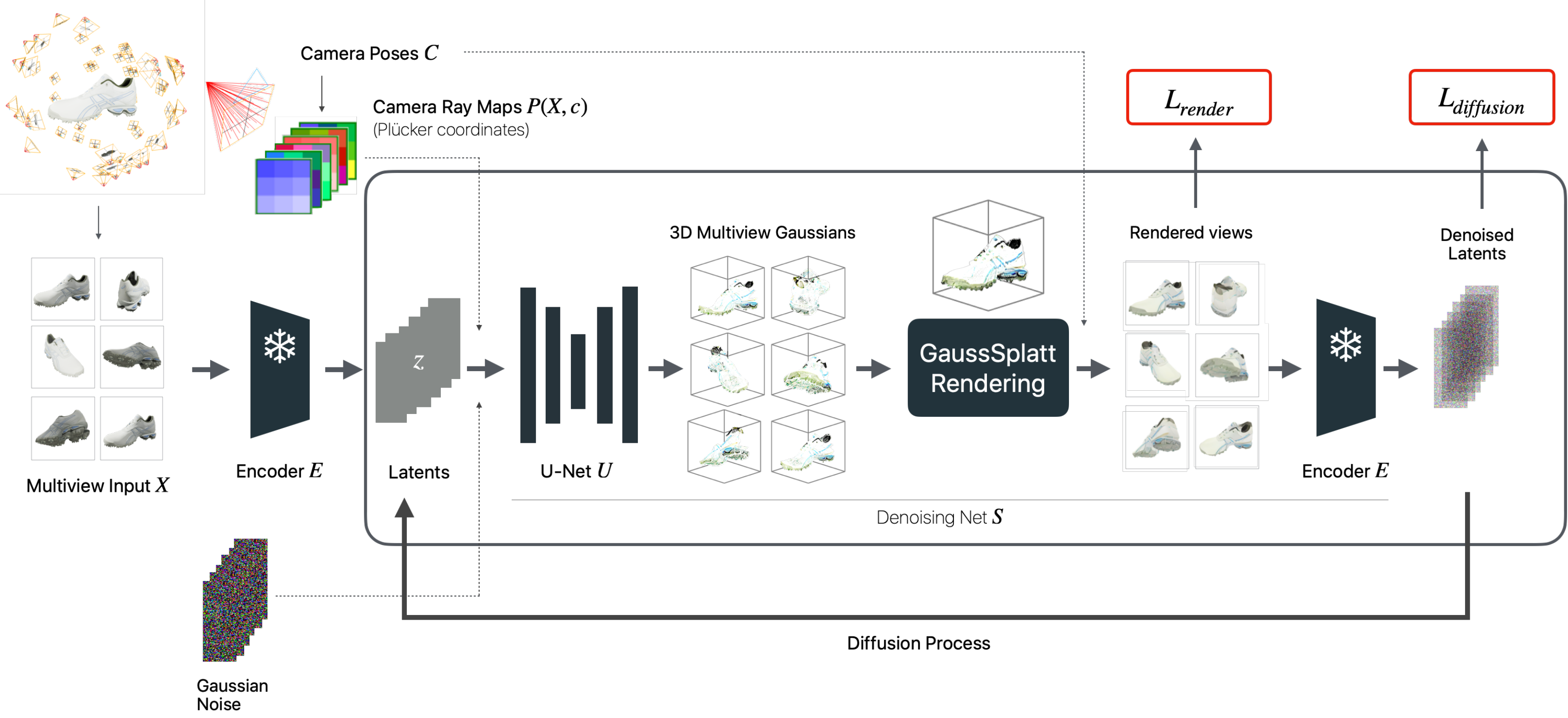}

    \caption{\work{}: single end-to-end training of an image pretrained and 3D aware diffusion model. During training time, we pass multiview input $X$ through our encoder to yield latents. Gaussian Noise is added to these latents and concatenated channel-wise with the Camera Ray Maps before being fed into the U-Net. The decoder outputs 3D multiview gaussians that are then used to render these multiview images as well as unseen views. The output renders are used to train our reconstruction model using $L_{render}$. Of the denoised output renders, we select the clean multiview images and encode them through our encoder to obtain denoised latents. These are used to train using  $L_{diffusion}$.}
    \label{fig:explainer}
\end{figure*}
\work \space introduces a novel training framework that unifies the expressive and robust 2D diffusion prior of latent diffusion models \cite{ho2020denoising} with the explicit 3D modeling capabilities of Gaussian Reconstructors~\cite{tang2024lgm, szymanowicz2024splatter, xu2024grm} in a single-shot manner.

Traditional Gaussian reconstruction models typically use a pixel-level U-Net as their backbone, making them incompatible with latent diffusion models (e.g., Stable Diffusion~\cite{rombach2022high}). To address this, we replace the backbone with a latent diffusion model and incorporate a Variational Auto-Encoder~\cite{kingma2013auto} to map images to and from the latent space using the encoder and decoder respectively. Since off-the-shelf latent diffusion models are optimized for image generation, we apply several modifications to adapt them for the reconstruction task. Within this work, the Reconstructor is jointly trained as a denoiser for multiview diffusion. 

Section \ref{3p} introduces the preliminaries, providing notation for multiview images and camera poses, as well as an overview of multiview diffusion, which addresses view consistency challenges in 3D generation. The model architecture is detailed in Section \ref{3a}, outlining the latent-space diffusion process, including the noisy encoder, 3D-aware denoising network, and Gaussian splatting for 3D model generation. Section \ref{3b} describes the training procedure, emphasizing the diffusion loss for denoising latents and the rendering loss for view reconstruction. Finally, the experimental setup, dataset details, and evaluation metrics are discussed in Section \ref{3d}, along with ablations that highlight the impact of pose conditioning, data mixture, and the number of input views. The overall training pipeline is summarized in Figure \ref{fig:explainer}.

\subsection{Preliminaries}
\label{3p}
\quad\textbf{Notation.} In the following we denote $X=(x^1, \ldots, x^v)$ as a set of images corresponding to $v$ views of a 3D object or scene. Each of these views is taken by a camera whose pose is encoded by $C=(c^1, \ldots, c^v)$.

\textbf{Multiview Diffusion.} Denoising Diffusion Probabilistic Models (DDPM) can be used to generate images by learning to denoise an image, i.e. by reversing a denoising process. During the forward process, Gaussian noise is incrementally added at each timestep $t \in \{1, \dots, T\}$, resulting in $q(x_t|x_0) = \mathcal{N}(x_{t}; \sqrt{\alpha_t}x_{0}, (1-\alpha_t)\mathcal{I})$, where $x_0=x$ is the original image and $\alpha_t$ represents a scalar value that controls the amount of noise added to the image at each timestep. After reparameterization, this represents the following forward process:
\[
x_t = \sqrt{\alpha_t}x_0 + \sqrt{1-\alpha_t}\epsilon_t
\]
where $\epsilon$ is sampled from $\mathcal{N}(0, \mathcal{I})$. In the reverse diffusion process, the model is trained to remove the noise applied during the forward pass, with $p_\theta(x_{t-1}|x_t)$, where $p$ represents a denoising neural network parameterized by $\theta$. 

Performing regular diffusion in generating or reconstructing 3D content poses significant challenges relating the view consistency across different renderings of the same object or scene. We address this issue by opting for multiview diffusion models instead, where we learn a joint probability $p_\theta$ of all object views $X$, conditioned on the view camera poses $C$. In practice, this means we can independently add noise to each input image according to the same noise schedule as in regular diffusion models:

\begin{equation}
X_t = ( \sqrt{\alpha_t} x_0^i + \sqrt{1 - \alpha_t} \epsilon^i_t \mid x_0^i \in X_0 )^v_{i=1} \end{equation}

\textbf{Gaussian Reconstructors.} Given multiple input images $X$ and camera poses $C$, the feed-forward reconstructor $R$ predicts $N$ Gaussians by directly regressing their parameters, conditioned on the camera pose encoding $P(x, c)$. The model outputs a set of 3D Gaussians parameterized as:
\[
\Theta = \{(X, Y, Z), \text{scale}, \text{color}, \text{opacity}, \text{orientation}\}_n^{N}
\]
where $N$ denotes the number of Gaussians . These parameters are subsequently rendered into 2D views using Gaussian Splatting, ensuring that the rendered outputs $\hat{X}$ align with the input images $X$. The spatially consistent representations of the Gaussians allow for high-quality view synthesis and efficient optimization across multiview settings.

In this work, we adopt and extend Gaussian Reconstructors as part of a unified multiview diffusion framework, leveraging their explicit 3D representation alongside the rich priors of latent diffusion models. This integration enables the generation of high-fidelity, spatially consistent multiview outputs and facilitates training through a combined rendering and diffusion loss.

\subsection{Model Architecture}
\label{3a}

Our model \( R \), referred to as reconstructor, takes as input multiple views of the same object/scene $X$ as well as camera poses $C$, one pose per view, and produces a 3D model. The model \( R \) performs denoising as a diffusion process in a latent space representing a 3D model. 

In particular, the model performs 3D generation via two steps. First, it encodes a set of model views $X$ with their camera poses into a latent space. This is accomplished via a \textit{noisy encoder} $E$. Second, it performs repeated denoising in this latent space via a \textit{3D-Aware Denoising Net} $S$. This network computes not only denoised latents, but also outputs an explicit 3D model and associated views. If we denote the part of the 3D-Aware Denoising Net that obtains the 3D model by $D$, then the final reconstructor model reads:
\begin{equation}
    \label{eq:full-model}
    R(X, C) = D(S^K(E(X), C), C)
\end{equation}
where we apply the denoising net $K$ times. Both the denoising net and the final 3D model computation require conditioning on a camera pose $C$.

We outline the details of the above networks next.

\textbf{Noisy Encoder.}  
The encoder maps each view $x^i$ independently into a feature map $z^i$, referred to as image latent. In more detail, the encoder is part of an autoencoder network that encodes and decodes images using the networks \( E_e \) (kept frozen and referred to as $E$ for simplicity) and \( E_d \) (trained and encapsulated as the last two layers of $U$), and is made of four downsampling and upsampling blocks each. Each down block in \( E_e \) consists of two ResNet layers, while each up block in \( E_d \) includes two cross-attention upsampling layers \cite{kingma}. The model is pretrained on the LAION dataset, optimized using a KL divergence loss, following Rombach et al. \cite{rombach2022high}. This means that for an image of dimension $w \times h \times 3$ the encoder produces latents of dimension $w/k \times h/k \times d$. In our implementation $k=8$ and $d=4$. After the feature map has been produced, the final latents are a noisy version of these features by adding uniform Gaussian noise $N(0, 1)$.

\textbf{3D-Aware Denoising Net.}
The denoising network $S$ operates on the full set of image latents $Z=(\ldots, z^i, \ldots)$, thus capturing a full 3D model. In particular, it attempts to denoise these image latents by explicitly constructing a 3D model represented as 3D Gaussians, rendering the same views and encoding them. Thus, it consists of three stages: first, it maps the latents to a set of Gaussians using a 12-block U-Net $U$; next, it renders the Gaussians using Gaussian Splatting; finally, the rendered images are encoded into latents using the above encoder $E$. Thus, the denoising network $S$ reads:
\begin{equation}\label{eq:denoiser}
 S(Z, C) = E(\textrm{GaussSplatt}(U(\textrm{Concat}(Z, C)), C))
\end{equation}

If we are concerned only with 3D model computations, then the model only uses the U-Net and it reads:
\begin{equation}
    \label{eq:3d-model-d}
    D(Z, C) = U(\textrm{Concat}(Z, C))
\end{equation}

For the architecture of the U-Net $U$ we closely follow \cite{rombach2022high} by having a convolutional network with 6 downsizing and 5 upsizing blocks (with the last two upsizing blocks coming from \( E_d \)), and corresponding skip connections across activations with the same spatial dimensions. Further, to leverage the prior knowledge from large collections of 2D images, we initialize the U-Net from a trained Latent Diffusion Model (LDM) – more specifically, the diffusers implementation of Stable Diffusion v2 \cite{rombach2022high, von-platen-etal-2022-diffusers}.

There are several noteworthy differences from the LDM implementation. First, we encode multiple latents in order to capture dependencies across them. Thus, we jointly encode the latents across all these views by arranging all $v$ latents into one single feature map of size $2w/k \times (v/2)h/k \times d$ and placing them into a grid of size $2 \times (v/2)$. Since the U-Net is a ConvNet, we can still initialize its weights from LDM while processing all latents jointly. One advantage in the above approach of conditioning the generative model on images is that we can flexibly change the number of images without having to modify or re-train our model.

Second, since we would like to output a 3D representation defined as a spatially arranged set of Gaussians, we reparameterize the last layer of the U-Net. In particular, we change its feature dimension to the number of parameters sufficient to describe a Gaussian. Since Gaussian Splatting requires color (represented in RGB space), the scale of the Gaussian, its 3D orientation (expressed in XYZ space), an opacity value, and spherical harmonics coefficient, this final layer produces $14$ features. Note that this layer is initialized randomly during training. This last layer of the U-Net then upscales the output to dimensions $h/2$ and $w/2$, resulting in approximately $92k$ Gaussians for a latent of size $256 \times 256$. Any 3D Gaussian with low opacity score, defined as less than $0.005$, is discarded from the final output.

Finally, the resulting Gaussians, representing a single object/scene, are rendered using Gaussian Splatting into $v$ views, each view $\hat{x}^i$ corresponding to the camera pose $c^i$. The output renderings also come with an alpha mask $M_{\alpha}$. These renderings are subsequently encoded separately into latents using $E$ and fed as input to the next denoising step.

\textbf{Camera Pose Encoding.} To encode the camera poses, we follow \citep{xu2023dmv3d} and employ Plücker coordinates as in \cite{zhang2024cameras}. In particular, we encode camera origin and orientation for each pixel in an image. The camera pose encoding $P(x, c)$ is a feature map of dimension $w \times h \times 6$ where for each pixel $(k, l)$ we encode the ray of that pixel in the world coordinate system. To do so, we capture both camera origin $o$ as well as the direction of that pixel $d_{k,l}$ using a vector cross product: $P(x, c)(k, l) = (d_{k, l}, o \times d_{k, l})$.

The downsized map is channel-wise concatenated to the latent $z$ and fed into the U-Net.

\subsection{Training}
\label{3b}
In order to obtain a full model we need to train the networks $S$ and $D$ from Eq.~(\ref{eq:full-model}). Note that $E$ is pre-trained as an encoder in a Variational Autoencoder setup and is kept frozen in our model. Since $\textrm{GaussSplatt}$ is a rendering procedure that has no trainable parameters, $D$ is the only trainable subnet inside $S$.

To obtain $D$ (and $S$ by proxy) we encourage two properties imposed by two losses. First, we want $S$ to approximate the reversal to a noise process in the 3D model latent space $Z=(z^1, \cdots z^v)$ in which $z^i = E(x^i)$:
\begin{equation}
Z_t = ( \sqrt{\alpha_t} E(x_0^i) + \sqrt{1 - \alpha_t} \epsilon_t \mid x_0^i \in X_0 )^v_{i=1}
\end{equation}
where we add noise $\epsilon_t$ with schedule $\alpha_t$ at noise step $t$.
We therefore introduce a diffusion loss that pushes the output of $S$ towards the above latents at every step $t$:
\begin{equation}\label{eq:denoise-loss}
    L_{\textrm{diff}}(t) = \lambda_3 ||S(z_t) - z_0||_2
\end{equation}

Second, we want to make sure that the intermediate 3D model yields renderings close to the input image views. To this end, we encourage the rendered views of the model, produced from noisy latents, to be close to the original clean model views:
\begin{equation}\label{eq:render-loss}
    L_{\textrm{render}}(t) = \lambda_1 ||\hat{X}_t - X_t||_{2} + \lambda_2 L_{\textrm{lpips}}(\hat{X}_t, X_0)
\end{equation}
where $\hat{X}_t=\textrm{GaussSplatt}(D(Z_t, C), C)$ are renderings of the denoised 3D model. We add both a pixel reconstruction loss as well as a perceptual distance loss based on LPIPS~\cite{zhang2018perceptual}.

The final loss is based on the diffusion denoising component from Eq.~(\ref{eq:denoise-loss}), as well as the rendering loss from Eq.~(\ref{eq:render-loss}):
\begin{equation}\label{eq:final-loss}
L(t)=\mathbb{E}_{X, C \sim X_{\textrm{full}}, C_{\textrm{full}}} (L_{\textrm{render}}(t) + L_{\textrm{diff}}(t))
\end{equation}

In this equation, $X_{full}$ and $C_{full}$ denote the complete set of images that we can sample from. Further details on the specific loss terms and pose conditioning strategies are provided in Section \ref{3d}. We observed that excluding supervision on unseen views can cause issues like collapse or flattening of 3D objects in the final outputs.

\textbf{Training Details.}
\label{3-training}
 As part of the training phase, we uniformly sample time step $t$ from $[1,1000]$ and add noise according to the cosine schedule. For this work, we fix $v=6$, similar to~\cite{shi2023zero123++}, where we render an object with the following azimuths: $\{30, 90, 150, 210, 270, 330\}$ and elevations: $\{20, -10, 20, -10, 20, -10\}$ respectively with a fixed camera radius of $1.5$ and fixed Field of View (FOV) of $50^{\circ}$. The first view becomes the clean conditioning signal during training. We then apply the loss objective as shown in Eq.~(\ref{eq:final-loss}). Empirically, we found that the L1-loss leads to slightly more stable results, especially when training on a difficult task (such as 3D reconstruction with background) or dirty data samples (e.g. motion blur, bright lighting). 

\begin{figure*}
    \centering
\includegraphics[width=0.95\linewidth]{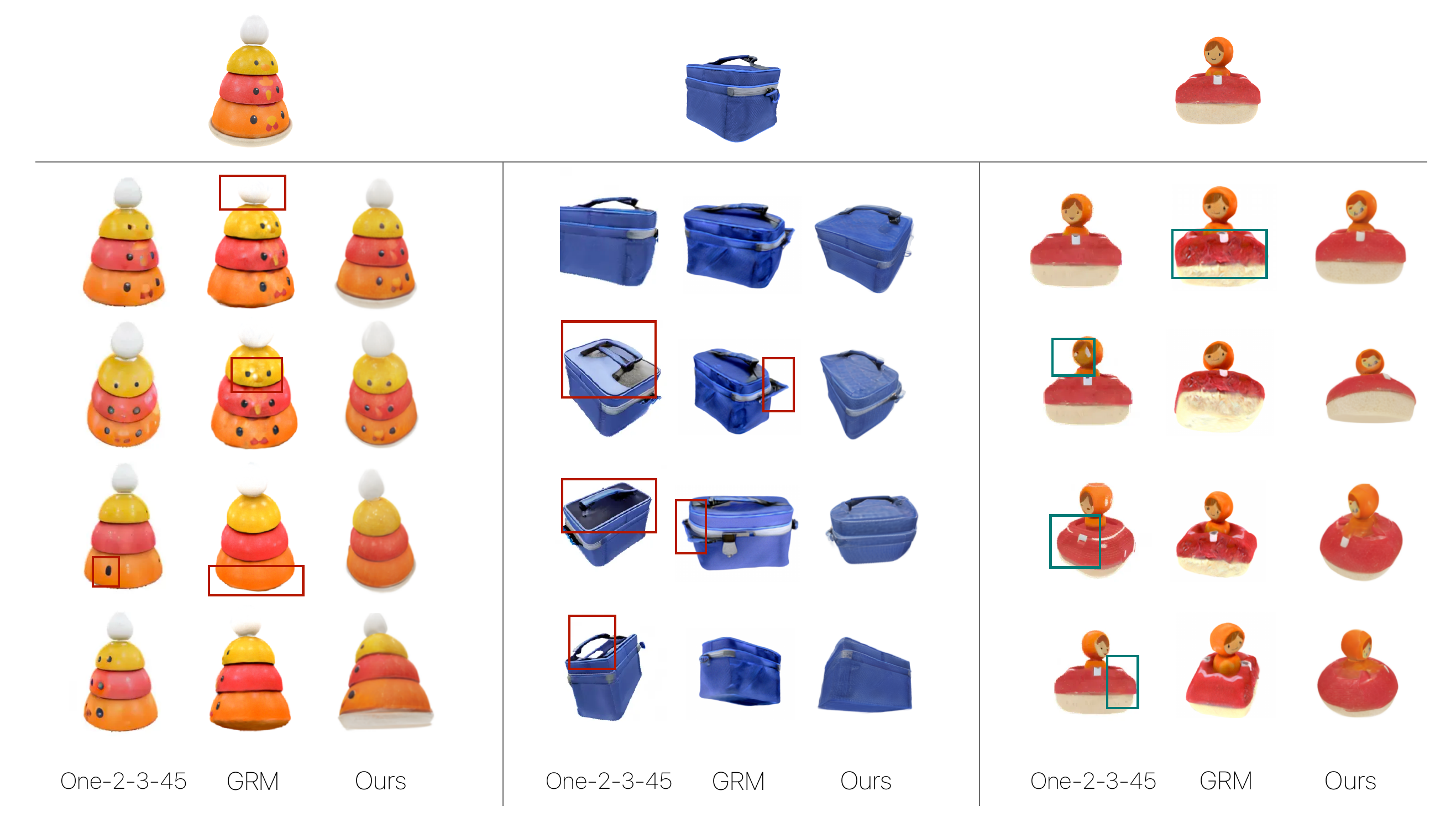}
    \caption{Qualitative comparisons of our results on Google Scanned Objects~\cite{downs2022google} with One-2-3-45~\cite{liu2024one} and GRM~\cite{xu2024grm}. Provided with a single input image (top row), we render four novel views for each of the methods. For One-2-3-45, we were unable to perfectly match the pose of the multiview images, so we display the image that is the closest approximation. From these results, it becomes clear that \work{} has strong photorealistic outputs (lighting, texture-wise), as well as a strong geometric prior.}
\label{fig:comparative results}
\end{figure*}

%% file: Paragraphs/4-experiments.tex
\section{Empirical Evaluation}\label{3d}

\subsection{Setup}

\textbf{Training Data.} We use two datasets:  Objaverse~\cite{deitke2023objaverse, deitke2024objaverse} and MVImgNet~\cite{yu2023mvimgnet}. Objaverse consists of approximately~$800K$ 3D objects with varying degrees of quality (e.g., missing textures, unconventional lighting conditions, broken meshes). Following \citet{yang2024hi3d}, we construct our dataset from the LVIS~\cite{gupta2019lvis} subset of Objaverse, consisting of approximately $44K$ object-centric, high-quality models. Additional sanity checks ensure that 3D objects with broken meshes or missing texture files are filtered out. To obtain views, we render 50 poses of each 3D model according to \citet{shi2023zero123++}. The strategy has been outlined in Section \ref{3-training}. The first 6 input views are fixed. The remaining 44 views are sampled uniformly at random from a sphere with radius 1.5, centered around the 3D object of interest. The output resolution is $(512 \times 512)$.

To extend the training of our model to real-world objects, we also leverage MVImgNet~\cite{yu2023mvimgnet}, a multiview image dataset that includes walkthroughs, poses/trajectories, and point clouds. This dataset contains around $200K$ assets, of which we utilized approximately $120$ after cleaning and extracting segmentation masks. 

\textbf{Test Data.}  
At test time, we use the Google Scanned Objects (GSO) dataset~\cite{downs2022google}, which contains 1000 real-world scanned objects. Following the same setup as GRM~\cite{xu2024grm}, we render 64 images for each object at four elevation angles, $\{10^{\circ}, 20^{\circ}, 30^{\circ}, 40^{\circ}\}$, and at evenly spaced azimuth angles. From this dataset, we select 250 objects for evaluation. For the input conditioning, we focus on samples with an elevation of $20^{\circ}$.

\textbf{Data Augmentation.} We apply two data augmentation strategies to improve the stability of the 3D reconstructor, especially given that the dataset includes synthetic samples: grid distortion and orbital camera jitter. Grid distortion mitigates subtle discrepancies across generated views by applying distortion to all views except for the first conditioning one. Orbital camera jitter introduces random noise to both input camera poses \( C \) and associated pose encodings \( P \), which are fed into the reconstructor \( R \), alongside slight rotations. This augmentation effectively captures the randomness in camera orientations typically observed in real-world data.

\vspace{-5pt}
\textbf{Implementation Details.} Our models are trained on 24 NVIDIA A100 GPUs, each with 80GB of RAM, for 100k iterations using bfloat16 precision. The training takes 5 to 7 days. The effective batch size is set to 96. In accordance with LRM and LGM conventions~\cite{xie2024lrm, tang2024lgm}, we transform all camera poses relative to the first input pose. The input images are of resolution $(256 \times 256)$, and the output renderings are of size $(512 \times 512)$. We use a learning rate of $2 \times 10^{-5}$. Both grid distortion and camera jitter are applied with a probability of 50\%. The DDPM noise scheduler with a cosine noise schedule is used for 1000 timesteps. During denoising, we use DDIM~\cite{song2020denoising} to accelerate the process, sampling in 50 inference steps.

\subsection{Evaluation and Analysis} 
\quad\textbf{Novel views of 3D Objects.} Following the GRM approach, we use the Google Scanned Objects dataset for evaluation. Specifically, for single-view based generation, models are evaluated using a view rendered at an elevation of $20^{\circ}$ as an input, with the remaining 63 renderings used for evaluation. We utilize a variety of metrics to assess quantitative performance, including PSNR~\cite{huynh2008scope}, LPIPS~\cite{zhang2018perceptual}, CLIP~\cite{radford2021learning}, and SSIM~\cite{wang2004image}. 

Quantitative results are presented in Table~\ref{table-1}. As we can see \work{} yields better performance than all other approaches, along both traditional metrics, such as PSNR, and perceptual metrics, such as LPIPS. In addition, in Fig.~\ref{fig:comparative results} we show comparative qualitative results showcasing that our approach yields novel views with either better semantics (correct lid color of the cooler bag), geometry (better reconstruction of the toy doll), or illumination (more homogenous and realistic surface color of the sides of the cooler bag and toy tower).

\begin{table}[t]
\centering
\label{tab:comparison}
\resizebox{\linewidth}{!}{%
\begin{tabular}{lccccc}
\hline
Method & PSNR $\uparrow$ & SSIM $\uparrow$ & LPIPS $\downarrow$ & CLIP $\uparrow$  \\
\hline
One-2-3-45 \cite{liu2024one} & 17.84 & 0.800 & 0.199 & 0.832 \\
Shap-E \cite{jun2023shap} & 15.45 & 0.772 & 0.297 & 0.854 \\
DreamGaussian \cite{tang2023dreamgaussian} & 19.19 & 0.811 & 0.171 & 0.862 \\
Wonder3D \cite{long2024wonder3d} & 17.29 & 0.815 & 0.240 & 0.871 \\
One-2-3-45++ \cite{liu2024one} & 17.79 & 0.819 & 0.219 & 0.886 \\
\hline
TriplaneGaussian \cite{zou2024triplane} & 16.81 & 0.797 & 0.257 & 0.840 \\
LGM \cite{tang2025lgm} & 16.90 & 0.819 & 0.235 & 0.855 \\
GRM \cite{xu2024grm} & 20.10 & 0.826 & 0.136 & \textbf{0.932} \\
\hline
% \textbf{\work \space (Ours)} & 17.89 & 0.831 & 0.141 & --- \\
\textbf{\work \space (ours)} & \textbf{20.38} & \textbf{0.842} & \textbf{0.109} & 0.921 \\
\hline
\end{tabular}}
\vspace{-5pt}
\caption{Single-Image to Multiview Reconstruction Results. Top approaches generate multiple object views followed by 3D reconstruction, while bottom approaches produce a 3D model directly.}
\label{table-1}
\end{table}

\begin{figure}
    \centering
    \vspace{-5pt}
    \includegraphics[width=1.0\linewidth]{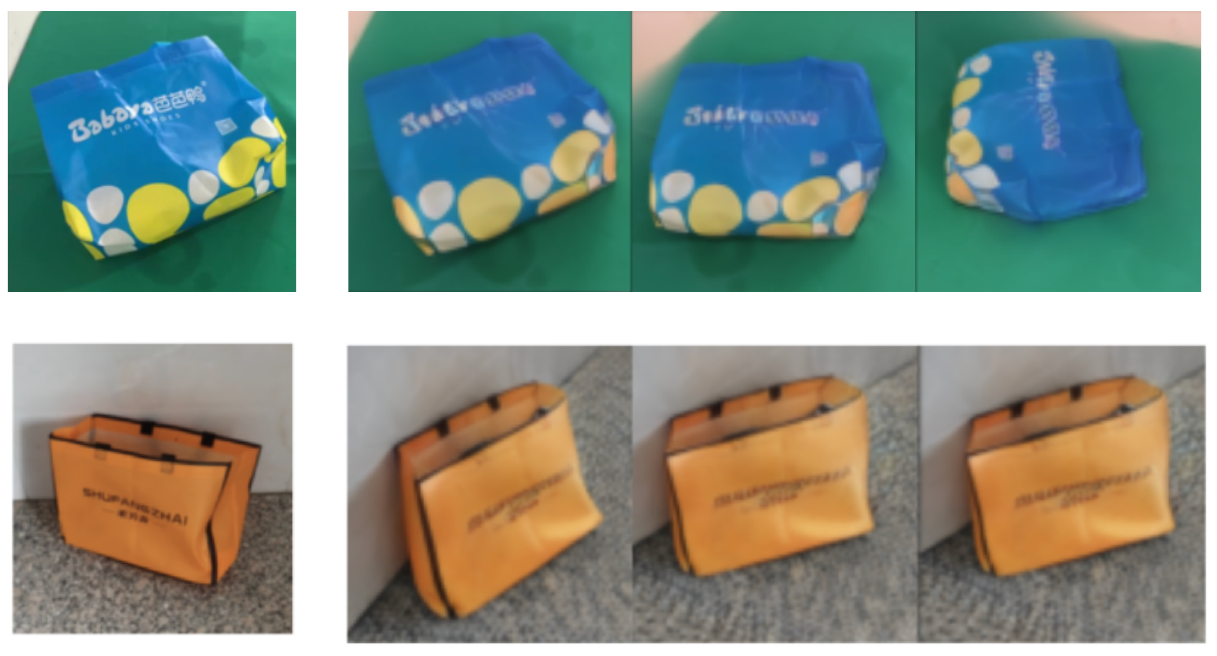}
    \caption{\work{} can be extended to real-world images, as shown in these objects in-the-wild (left) and the corresponding generated novel views (right).}
    \label{fig:scene-generation}
\vspace{-5pt}
\end{figure}

\textbf{Novel Views of Scenes.} In addition to applying \work{} on single-object images and reconstructing only the object, we also show that our method can generate novel views of objects captured in the wild. The challenge of this task is in the length of the video trajectory as well as the scene complexity, i.e. the presence of a background. For this application, we generate views over a set of $v$ frames that are evenly spaced out between the first and last pose in our trajectory. Since an established benchmark for this task does not exist, we visualize a couple of examples in Fig.~\ref{fig:scene-generation}. We see that \work{} is capable of generating novel views with realistic geometry for both an object and its background, including a realistic illumination.

\begin{table}[t]
\centering
\label{tab:ablate}
\resizebox{\linewidth}{!}{%
\begin{tabular}{lcccc}
\hline
Data & Views& Pose cond. & LPIPS $\downarrow$ & PSNR $\uparrow$ \\
\hline
\multirow{2}{*}{Objaverse} & 6 & Yes & 0.143 & 18.95 \\
 & 6 & No & 0.155 & 18.45 \\
Objaverse $+$ & 6 & Yes & 0.109 & 20.38 \\
MVImgNet & 4 & Yes & 0.123 & 19.55 \\
\hline
\end{tabular}}

\caption{Ablation evaluation by modifying training data, number of conditioning views during training, and presence or absence of pose conditioning. Metrics computed on GSO.}
\vspace{-10pt}
\label{tab:ablations}
\end{table}

\textbf{Ablations.} In order to better understand \work{}, we perform ablations with respect to training data mix, the number of conditioning views during training, as well as presence of pose conditioning (see Table \ref{tab:ablations}).

In particular, we experiment with training on Objaverse-only or a mix of Objaverse and MVImgNet with a mixing ratio of 2:1. The larger mix leads to an improvement of $1.5$ in PSNR and $0.34$ in LPIPS, demonstrating the importance of real world data.

Furthermore, we train a model on 4 views instead of 6, using the first 4 poses of our pre-defined poses, to assess the importance of more views at train time. Using fewer views leads to a marginal drop in performance.

Lastly, we also ablated the importance of including pose conditioning as an input to the 3D Denoising U-Net. Without pose conditioning we use 6 predefined poses for all examples during training. The difference is less than $0.5 \%$ for both PSNR and LPIPS values. This suggests that the camera views embeddings do not provide a significant contribution, but might help stabilize the model and yield a minor performance boost.

%% file: Paragraphs/5-conclusion.tex
\vspace{-0.2cm}
\section{Conclusion}
\vspace{-0.2cm}
\work~combines the strong 2D diffusion prior of Latent Diffusion Models with the explicit 3D representations of Gaussian-based Sparse View Reconstruction models. In addition, we introduce how to train the 3D aware diffusion model in a single-shot fashion. Evaluations show state-of-the-art performance as well as photorealistic and geometrically-correct 3D outputs. 

There remain several areas of opportunity. \work~ fundamentally works by full coverage of said 3D object(s) that are either implicitly or explicitly parameterized. It remains unclear how much our approach generalizes to settings with larger intervals, non object-centric (i.e. scene data).

%% file: Paragraphs/6-appendix.tex
\ifincludeappendix
\appendix
\clearpage
\setcounter{page}{1}
\maketitlesupplementary

\newpage
\section{Supplementary Material}

\subsection{Training Data Curation}\label{appendix
}

\textbf{Rendering} To ensure consistency across all training renders, we use Blender~\cite{blender} to normalize the assets to be within the bounding box with coordinates in the $(-1,1)$ range. Our lighting setup is optimized to balance ambient and directional lighting, ensuring consistent shading across objects. Each input view's orientation contains an azimuth that is sampled uniformly at random and fixed elevation of $0^{\circ}$. The remaining views are rendered as outlined in Section 4.1 at a resolution of $(320 \times 320)$ and a fixed camera radius of $1.5$. 

\textbf{Selection} We train on a subset of the LVIS dataset of Objaverse~\cite{objaverse, objaverseXL} which is a high quality subset containing object classes for 3D objects aligned with ImageNet~\cite{deng2009imagenet}. During curation, additional filtering steps remove instances with missing or incomplete textures to ensure high-quality input data across the board. We also remove samples that contain either too little or too pronounced background.

\subsection{More Results}\label{appendix:data}

\textbf{Objects} Figure \ref{appendix:suppl_qual} presents additional qualitative results produced by \work, illustrating the input view alongside six generated multiview outputs. The input images include a combination of test samples from the Google Scanned Objects (GSO) dataset and publicly available internet sources. The results demonstrate that our model generates photorealistic outputs, as evident in the texture details of the jacket. Furthermore, the fidelity of fine details is highlighted in examples such as the intricacies of the frog sample. Our model exhibits a strong spatial understanding of 3D structures and geometry, as demonstrated in the the toy bear, cow, and woven basket. It also effectively captures nuanced details, such as the texture present in the foliage of the tree.

\textbf{Objects in the Wild}\label{appendix:qualdata} In addition to reconstructing isolated objects, we also demonstrate \work's ability to regenerate 3D objects situated within complex scenes, even when all viewpoints of the objects are not accessible. To address this challenge, we train our model on the MVImgNet dataset~\cite{yu2023mvimgnet} and leverage the camera conditioning module as outlined in Section 3.3. During training, we evenly divide the walk-through samples into six segments, ensuring that the first frame of each segment is included as part of the multiview input. These multiview inputs are shown in Figure \ref{suppl:scene-generation-mv}.

\begin{figure}
    \centering
    \includegraphics[width=1.0\linewidth]{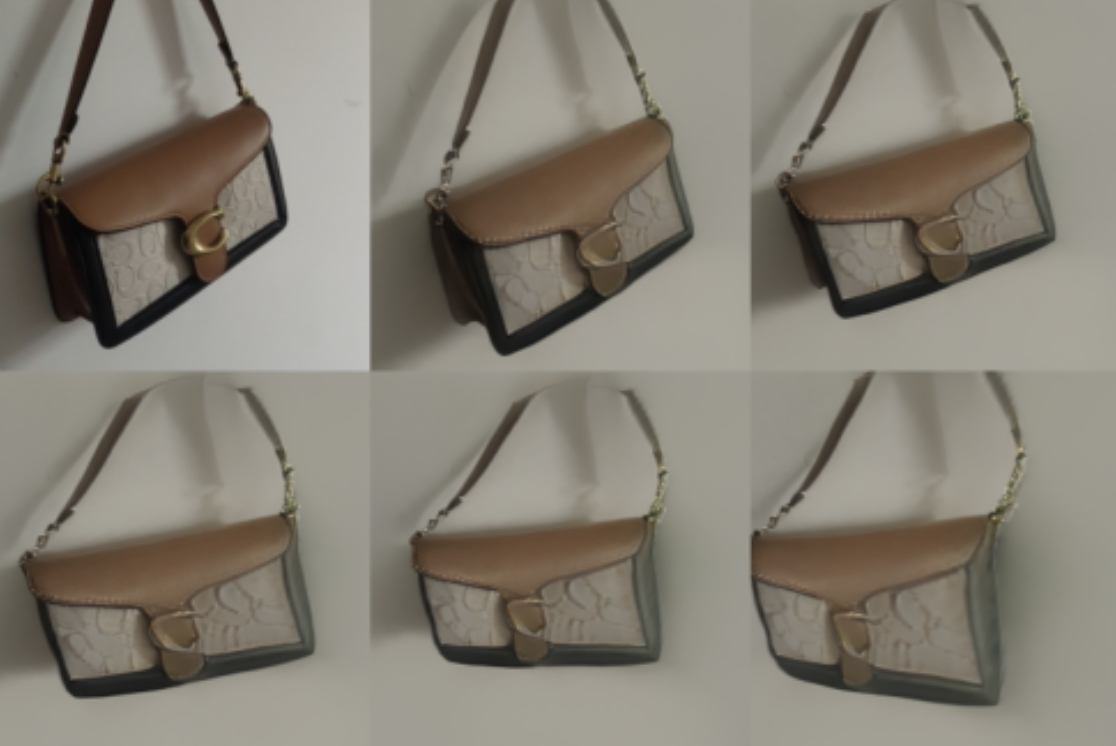}
    \vspace{5pt}
    \includegraphics[width=1.0\linewidth]{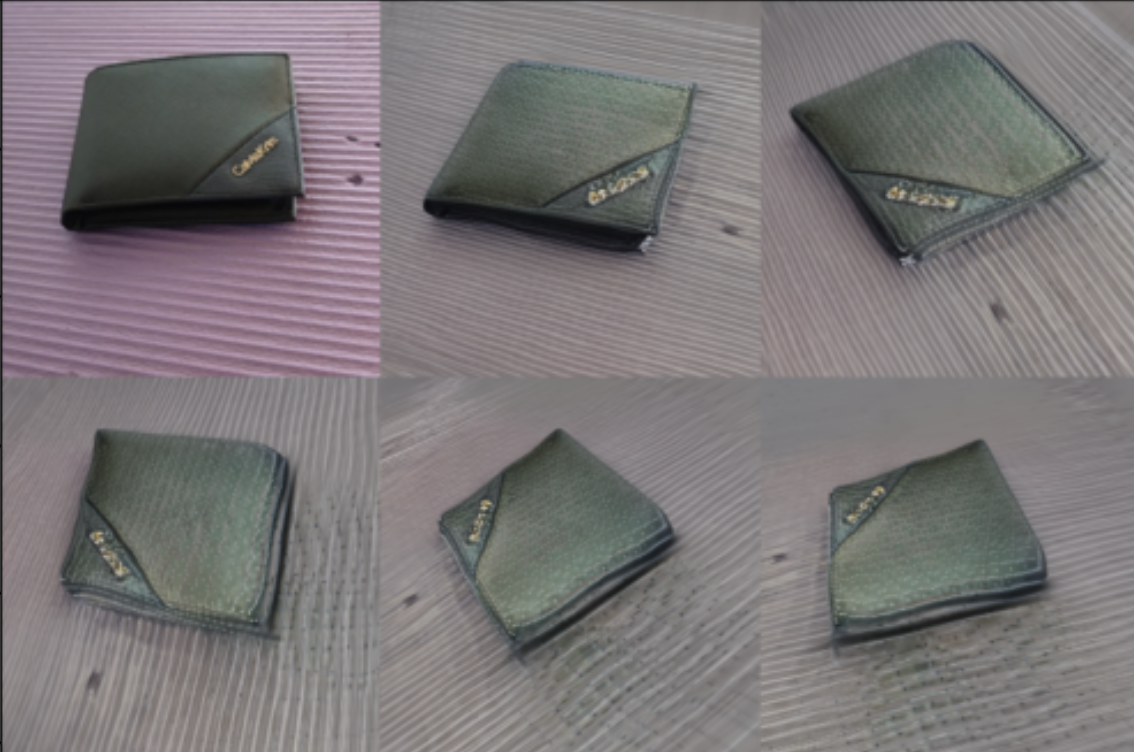}
    \caption{Additional qualitative results for Objects in the Wild generation. In the top left of both images, the input views are shown. The remainder of the images are the multiview image outputs for reconstructing the trajectory containing the bag and wallet respectively.}
    \label{suppl:scene-generation-mv}
\vspace{-15pt}
\end{figure}

\subsection{Limitations}\label{appendix:limitations}

Informed by our failure cases, we acknowledge that there remain some limitations to our model's performance. The main one that we battled with was the widespread 'lightening effect' of Gaussian Splatting~\cite{kerbl20233d}, which makes images slightly brighter and less saturated due to a set of uncertain Gaussians. Additionally, as mentioned in \cite{tang2024lgm}, our model also struggles in capturing high-frequency textural/geometrical details, as well as straight structures; this could potentially be mitigated by increasing the total resolution of the Gaussians.

\clearpage

\begin{figure*}
    \centering
    \includegraphics[width=0.95\linewidth]{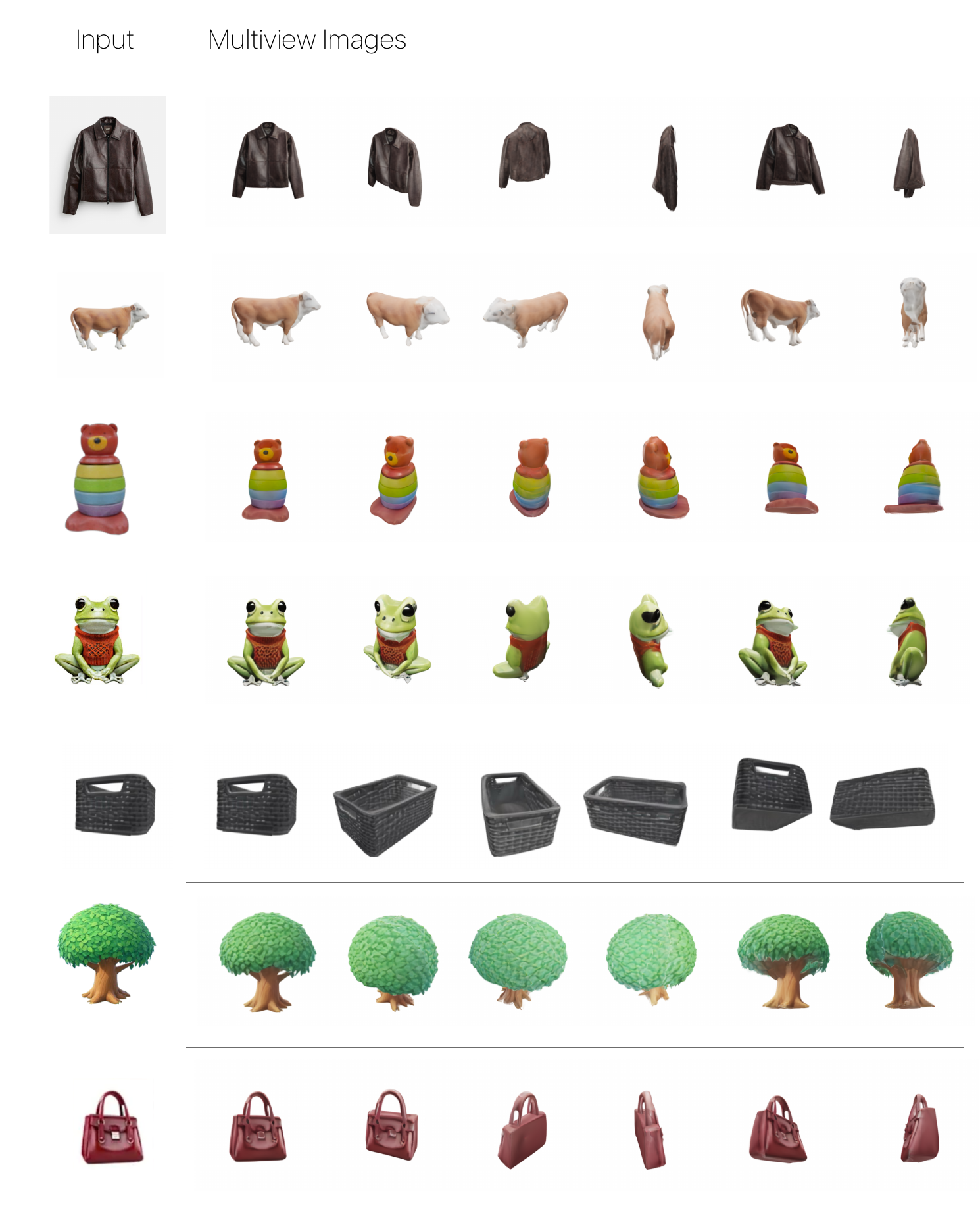}
    \caption{Additional Qualitative Analysis showcasing the 6 surrounding multiview images generated by \work. In this example, the input images are either testing images from GSO or have been sourced from other public resources.}
    \label{appendix:suppl_qual}
\end{figure*}

\clearpage

\else
\newcommand{\dummyappendix}{}
\expandafter\gdef\csname r@sec:appendix-section\endcsname{Appendix}
\fi